\documentclass[12pt]{article}
\ifx\pdfoutput\undefined
\usepackage[dvips]{graphicx}
\else
\usepackage[pdftex]{graphicx}
\usepackage{type1cm}
\fi
\newcounter{sec}
\newcommand{\mysection}[1]{\section{#1}
    \hspace{0.8cm}\setcounter{equation}{0}}

\newlength{\dummysp}
\newcommand{\diag}{\mathop{{\hbox{diag} \, }}\nolimits}

\newcommand{\beqa}{\begin{eqnarray}}
\newcommand{\eeqa}{\end{eqnarray}}
\newcommand{\nnn}{ \nonumber \\ }

\newcommand{\p}{{\partial}}

\newcommand{\chib}{{\bar \chi}}

\newcommand{\s}{{\sigma}}

\newcommand{\bigvev}[1]{{\left\langle #1 \right\rangle}}

\newcommand{\gappeq}{\mathrel{\rlap {\raise.5ex\hbox{$>$}}
{\lower.5ex\hbox{$\sim$}}}}
\newcommand{\lappeq}{\mathrel{\rlap{\raise.5ex\hbox{$<$}}
{\lower.5ex\hbox{$\sim$}}}}
\newcommand{\myref}[1]{(\ref{#1})}

\newcommand{\ben}{\begin{enumerate}}
\newcommand{\een}{\end{enumerate}}

\newcommand{\etab}{{\bar \eta}}
\newcommand{\sqtw}{\sqrt{2}}

\newcommand{\sbar}{{\bar \s}}

\newcommand{\alphadot}{{\dot \alpha}}
\newcommand{\adot}{\alphadot}
\newcommand{\saat}{{\s_{\alpha \adot}^3}}
\newcommand{\hc}{{\rm h.c.}}

\newcommand{\psib}{{\bar \psi}}

\newcommand{\ddd}{\nnn &&}

\newcommand{\vs}{|v|^2}
\newcommand{\va}{|v|}

\newcommand{\lb}{{\bar \lambda}}
\newcommand{\vp}{\varphi}
\newcommand{\jp}{{j'}}
\def\[{\left [}
\def\]{\right ]}
\def\({\left (}
\def\){\right )}
\newcommand{\beq}{\begin{eqnarray}}
\newcommand{\eeq}{\end{eqnarray}}

\begin{document}

\baselineskip=18pt

\begin{titlepage}

\begin{center}

\hfill {\tt hep-th/0301048}\\

\vskip 1.2in

{\LARGE \bf Deconstruction, Lattice Supersymmetry,
\vspace{.3cm}

Anomalies and
Branes}

\vskip .5in

{\bf
      Joel Giedt$^a$,  Erich Poppitz$^a$, and   Moshe Rozali$^b$}

\vskip 0.2in

$^a${\em Department of Physics, University of Toronto

Toronto, ON, M5S 1A7, Canada} \vskip 0.1in

$^b${\em Department of Physics and Astronomy, University of
British Columbia

Vancouver, BC, V6T 1Z1, Canada} \vskip 0.1in

{\tt giedt@physics.utoronto.ca, poppitz@physics.utoronto.ca,
rozali@physics.ubc.ca}
\end{center}

\vskip .5in
\begin{abstract}
We study the realization of anomalous Ward identities in
deconstructed (latticized) supersymmetric theories. In a
deconstructed four-dimensional theory with $N=2$ supersymmetry, we
show that the chiral symmetries only appear  in the infrared  and
that the anomaly is reproduced in the usual framework of  lattice
perturbation theory with Wilson fermions. We then realize the
theory on the world-volume of fractional D-branes on an orbifold.
In this brane realization, we show how deconstructed theory
anomalies can be computed via classical supergravity. Our methods
and observations are more generally applicable to
deconstructed/latticized supersymmetric theories in various
dimensions.
\end{abstract}
\end{titlepage}

\newpage

\mysection{Introduction and Summary}

Deconstruction was originally proposed as a gauge-invariant
regularization and ultraviolet completion of five- and
six-dimensional gauge theories in terms of a four-dimensional
theory  with a particular product gauge group, a ``moose"
theory \cite{Arkani-Hamed:2001ca}. (Theories with a
product of $N$ gauge groups and matter fields
linking the groups were considered earlier in
\cite{Georgi:au} and \cite{Halpern:1975yj}, albeit with a very different
motivation and interpretation.)

These  ``moose," or ``quiver,"  product-group   theories appear
naturally  in string theory as  world-volume theories on branes  at
orbifold singularities \cite{Douglas:1996sw}. The relation of
deconstruction to string theory has led to some interesting
proposals. One is the idea  that a  four dimensional ``moose"
theory may provide, in an appropriate limit, a definition of the
ill-understood $(0,2)$ supersymmetric six-dimensional theory
\cite{Arkani-Hamed:2001ie}.


Another proposal is based on the observation that the brane
realization of deconstruction can  be used to give a definition of
three- and four-dimensional supersymmetric gauge theories in terms
of one-  or zero-dimensional product-group  theories. These
one- (zero-) dimensional theories can provide a spatial
\cite{Kaplan:2002wv}  (or Euclidean \cite{cohenprivate}) lattice
formulation of supersymmetric theories; see also
\cite{Rothstein:2001tu}. The lattice theory explicitly
preserves part of the supersymmetry of the
continuum theory. The explicit  supersymmetry, along with  other
symmetries of the deconstructed theory forbids relevant operators that
break
the continuum supersymmetry, and ensures  that the
  enhanced continuum supersymmetry is achieved without
fine tuning \cite{Kaplan:2002wv}.

This development  leads to the hope that deconstruction may
ultimately be useful to study  strong-coupling supersymmetric gauge
dynamics in three or four dimensions via numerical simulations;
for a review of the progress in this direction, see
\cite{Feo:2002yi}. This is important, because many interesting aspects
of supersymmetric gauge
dynamics, relevant for various applications, are not accessible by
the tools---holomorphy and symmetries---normally used to obtain
exact results.

An important tool in the study of   nonperturbative gauge dynamics
  is the understanding of
chiral symmetries and   anomalies (both gauge and 't Hooft).
In this paper, we investigate the chiral symmetries and their
anomalies in the deconstructed (lattice) formulation of
supersymmetric theories. We note that some related issues---the parity
anomaly and the Chern-Simons term---in the deconstruction of
five-dimensional theories have been addressed before
\cite{Csaki:2001zx,Skiba:2002nx}; see also \cite{Constable:2002vt}.
Here we study the deconstruction of four-dimensional theories,
with continuous global chiral symmetries
with gauge or 't Hooft anomalies. The deconstructed supersymmetric
theories we study
  are vector-like.  Applying deconstruction to chiral gauge theories can
be done in principle, but is expected
  to face
  the usual problems of putting chiral fermions on the lattice. We
  hope to return to this issue in the near future.

We consider a simple example: the four-dimensional $N=2$
supersymmetric pure Yang-Mills theory on a one-dimensional spatial
lattice, i.e. its deconstruction  in terms of a three-dimensional
product-group theory. This setup is relatively simple, allowing us
to look at chiral symmetries and anomalies from different points
of view, while keeping the expressions reasonably tractable; yet, it
contains features generic enough to be shared by other
deconstructed/lattice supersymmetric theories (e.g. 4d $N=4$, or
3d $N=4$,  SYM on a spatial lattice \cite{Kaplan:2002wv};
the deconstruction of higher-dimensional supersymmetric theories,
etc.). In fact, all our findings can be generalized to these
examples.

{\flushleft B}elow, we give a summary of our results and point to
the relevant sections of the paper:
\begin{enumerate}
\item{
We find that the chiral symmetry of the theory is not explicit at
finite lattice spacing and is broken by a Wilson term, with
coefficient fixed by the (super-)symmetries of the deconstructed theory.
Thus the chiral symmetry appears only as an accidental symmetry in
the continuum limit. The lattice action is worked out in section
2.1, and the equivalence to the standard Wilson fermion
formulation is shown in section 2.2.}
\item{The anomaly of the chiral $U(1)_R$ symmetry can be understood
using lattice perturbation theory, as in
the classic work of Karsten and Smit \cite{KS81}. A point worth
noticing is that the large value of the Wilson term coefficient
precludes the interpretation of the anomaly as due to heavy
doublers acting as  4d Pauli-Villars regulators in the continuum
limit, yet the correct anomalous Ward identity results. This is
discussed in section 2.2.}
\item{A ``modern" view on the chiral anomaly in the deconstructed theory
results from considering its fractional brane realization. This brane
realization is explained in sections 3.1, 3.2. In section 3.1,  we also
review how the $U(1)_R$ anomaly  in the continuum Seiberg-Witten
theory is understood from the corresponding classical
supergravity.}
\item{We show, in  section 3.3,  how one can compute the anomaly in the
lattice version of the theory using fractional branes. We also find the
expression for the running coupling and discuss the continuum limit. In
section 3.4, we give a geometric interpretation of the R-symmetry
violating Wilson term.}
\end{enumerate}

{\flushleft To} summarize, we have shown that even though the
chiral symmetries are explicitly broken in the
deconstructed/latticized supersymmetric theories, the correct
anomalous Ward identities are obtained in the continuum limit. We show
this  from two different  points of view: lattice perturbation
theory and the brane realization of   deconstruction. Our
observations are more generally valid than the model  we
considered: for example, fractional branes can  be used to study
the gauge coupling running and the Chern-Simons coefficient in
deconstructed $N=1$ 5d theories.

We do not yet have a detailed understanding of the anomaly in the
deconstructed theory as    charge nonconservation in a nontrivial
gauge field background. This question is currently under
investigation; it requires a better understanding of the map between
nonperturbative effects in the deconstructed and continuum
theories.

\mysection{Global Chiral Symmetries and Anomalies in the
Deconstructed Seiberg-Witten Theory}
  In
this section, we study the global chiral symmetries of the
classical action of the
  deconstructed $N=2$ 4d Yang-Mills theory. We show that the global
chiral $SU(2)_R \times U(1)_R$ symmetry of the continuum
theory is explicitly broken by the lattice action.  The anomaly of
the $U(1)_R$ symmetry is reproduced, in the continuum limit, by
lattice perturbation theory, similar to \cite{KS81}.

\subsection{The Deconstructed Theory Action and Symmetries}

We begin by writing down the action of the deconstructed theory,
starting from the superspace expression. The theory whose lattice
action we
  want to explicitly study is the deconstruction of a 4d $N=2$ Yang-Mills
theory with gauge group
$SU(k)$. Its deconstruction is the dimensional reduction,  from 4d
to 3d, of the theory considered in \cite{Csaki:2001em}.  It is
convenient to retain four-dimensional superfield notation
throughout, while simply ``forgetting" the dependence on the
coordinate $x^3$.

The deconstructed theory is a $SU(k)^N$   quiver gauge
theory\footnote{For simplicity, we  ignore the $U(1)$ factors
present in the lattice models of \cite{Kaplan:2002wv}, as well as in the
D-brane construction,  which have
a $U(k)^N$ gauge group. We
only note that the strength of the  $U(1)$ D-term potential might
be used to  stabilize the relative fluctuations of the $Q_i$ vevs
(\ref{links}), i.e. the difference between lattice spacings on
neighboring links (but not of the overall zero mode; this issue is
not relevant for our 3d to 4d deconstruction, so we do not address
it here).}  with $N=2$ 3d supersymmetry  (four real supercharges).
The gauge fields are in vector supermultiplets $V_i$, while the
chiral matter superfields $Q_i$  transform as a fundamental under
the $i$-th gauge group and antifundamental under the $i+1$-th
gauge group.
In other words $Q_i \rightarrow e^{\Lambda_i } Q_i
e^{- \Lambda_{i+1}}$, $e^{V_i} \rightarrow e^{- \Lambda_i^\dagger}
e^{V_i} e^{- \Lambda_i}$ under gauge transformations.
The $N+1$-th gauge group is identified with the first, so
that a closed circular ``moose" is obtained.
The superspace lagrangian is described by the
tree-level K\" ahler potential:
\beq
\label{kahler}
K~ = ~{1 \over
g_3^2} ~\sum\limits_{i=1}^N {\rm tr} ~Q_i^\dagger ~e^{V_i}~ Q_i~
e^{- V_{i+1}}~.
\eeq
The $F$-term part of the tree-level
lagrangian\footnote{In the superrenormalizable $N=2$ 3d supersymmetric
theory  (\ref{kahler}), (\ref{super}), there are no divergent
corrections to the tree level lagrangian. Finite loop effects are
discussed in section 3.3.}  arises from the gauge kinetic terms: \beq
\label{super}
W ~=~ {1 \over 2  g_3^2}~ ~\sum\limits_{i=1}^N {\rm tr} ~W^\alpha
(V_i) ~W_\alpha(V_i)~. \eeq The superpotential is forbidden by the
$U(1)_R$ symmetry of the theory, part of the $U(1)^N \times U(1)_R
\times Z_N$ global symmetry of the model (here $U(1)_i$ is the
global symmetry under which the $Q_i$ have unit charge and $Z_N$
is the ``rotational" symmetry of the moose; the theory also
preserves 3d parity). We keep the canonical
dimensions of all fields appropriate to 4d, so $g_3^2$ has dimension of
mass. We
use generators in the fundamental, so that tr $T^a T^b =
\delta^{ab}/2$ and $[T^a, T^b]= i f^{abc} T^c$. Further, we also
need $g^{abc} =$ tr($T^a T^b T^c) = {1\over 4} i f^{abc} + {1\over
2} d^{abc}$, with $d^{abc} = {\rm tr}\ T^a \{ T^b, T^c\}$.

Below, we work in Wess-Zumino gauge. We let $m = 0, 1,2$ and
$\mu = 0, 1, 2,3$. The $A_3$ component of the 4d vector
boson is the real adjoint scalar in the 3d $N=2$ vector
multiplet. Even though in 3d we can use the charge conjugation
matrix $\sigma^3_{\alpha \dot\alpha}$ to convert dotted into
undotted indices and write the entire lagrangian in terms of one
type of indices, we will keep the dotted and undotted spinor
indices as this simplifies the consideration of the 4d limit.

As was shown in \cite{Csaki:2001em} by considering the mass
spectrum, the theory  (\ref{kahler}), (\ref{super}) describes,
along the Higgs branch  $\langle Q_i \rangle = v$ and in the
large-$N$ limit,
  a supersymmetric field theory with gauge group $SU(k)$ in one higher
dimension and with
   twice the   amount of  supersymmetry, in our case---an  $N=2$ 4d
supersymmetric theory.

Our aim here is to study the interactions and the symmetries of
the theory.  To describe the relevant  (i.e. the ones charged under
$SU(k)^N$, keeping the
vevs frozen, see footnote 1) fluctuations of the link
fields $Q_i$, we define: \beq \label{links} Q_i =
v~(1 + q_i^a T^a)~. \eeq
It is straightforward to isolate  the fermion bilinear
terms corresponding to (\ref{kahler}) and (\ref{super}): \beq
\label{fermion1} L_{ferm.} &= & {|v|^2 \over g_3^2} \sum\limits_{j
= 1}^N  \left[ -{i\over 2} \bar\psi_j^a \bar\sigma^m  \partial_m
\psi_j^a + {i \over 2 \sqrt{2}} ~ \psi_j^a ( \lambda_j^a -
\lambda_{j+1}^a) - {i \over 2\sqrt{2}}  ~\bar\psi_j^a (
\bar\lambda_j^a - \bar\lambda_{j+1}^a)
\right. \nonumber \\
&& \left. - {1\over 2} ~g^{abc} ~ \bar\psi_j^a \bar\sigma^\mu
\psi_j^b~ A^{j+1,c}_\mu +  {1\over 2} ~g^{abc} ~ \bar\psi_j^a
\bar\sigma^\mu \psi_j^c~ A^{j, b}_\mu + \right. \nnn && \left. +
{i\over  \sqrt{2}}~ g^{abc} \left( q^{b}_j ~\bar\lambda_{j+1}^c
\bar\psi_j^a -
    q^{*,a}_j~  \lambda_{j+1}^c \psi_j^b\right)
-    {i\over  \sqrt{2}}~ g^{abc} \left( q^{c}_j ~\bar\lambda_{j}^b
\bar\psi_j^a -
    q^{*,a}_j  ~\lambda_{j}^b \psi_j^c \right)
\right]  \nonumber  \\
& & + {1 \over g_3^2}~\sum\limits_{j = 1}^N \left[ - i
\bar\lambda_j^a \bar\sigma^m  \partial_m \lambda_j^a +  {i\over 2}
f^{abc} \bar\lambda_j^a \bar\sigma^\mu   \lambda_j^c A^{j,b}_\mu
\right]  ~.
\eeq

The expression (\ref{fermion1}) will be useful in
section 2.2, when comparing to the Wilson fermion action.
However, we will first
consider the fermion lagrangian (\ref{fermion1}) in momentum space
along the latticized direction.   Diagonalizing the kinetic terms
helps to reveal  the 4d nature of the theory and the
enhanced supersymmetry in the continuum limit, as well as to
elucidate the finite-$N$ symmetries of the fermion interactions.

To write down the Fourier space action,  we will proceed in two
steps, skipping the intermediate results, but explicitly giving
the relevant field redefinitions. First, we perform the following Fourier
transformations in (\ref{fermion1}), assuming odd $N$ for
definiteness:
\beq \lambda_j &=& {1\over \sqrt{N}}
~\sum\limits_{k = -{N-1\over 2}}^{N-1\over 2} \omega_N^{jk
-{k\over 2}}~ \lambda_k, \qquad \psi_j  =  {\sqrt{2} \over
\sqrt{N} |v|} ~\sum\limits_{k =  -{N-1\over 2}}^{N-1 \over 2}
\omega_N^{-jk}~ \psi_k~,
\nnn
A_{\mu}^{k,a} &=& {1\over \sqrt{N} } ~ \sum\limits_{j= 1}^N
A_{\mu}^{j,a}~ \omega_N^{j k-{k \over 2}}~, \qquad
q_k^a = {|v| \over \sqrt{2 N} } ~ \sum\limits_{j = 1}^N q_j^{a}~ \omega_N^{j k}~,
\nnn q_k^{* a} &=& {|v| \over \sqrt{2 N} } ~ \sum\limits_{j = 1}^N
q_j^{* a}~ \omega_N^{- j k} = (q_k^a)^*, \eeq
where $\omega_N =
e^{2 \pi i\over N}$, and the complex conjugate transformation for
the dotted fields. Note also the reality condition $A^{-k} = (A^{k})^*$.
It is easy to verify that   terms  proportional to $d^{abc}$
vanish in the continuum limit. We will thus (for simplicity only)
further ignore them even for finite $N$, concentrating, e.g., on
the $SP(n)$ case.
Finally, note that the
lagrangian is invariant under 3d parity, with $\psi(-x_2) =  i
\sigma^3 \bar\sigma^2 \psi(x)$, $\lambda(-x_2) = - i \sigma^3
\bar\sigma^2 \lambda(x_2)$,  $A_{2 (3)}(-x_2) = - A_{2 (3)}(x_2)$,
and $A_{0 (1)}(-x_2) = A_{0 (1)}(x_2)$.  Invariance of the mass
term requires opposite intrinsic parities of the gauginos and
bifundamental fermions.

The second step in revealing the   four-dimensional Lorentz
invariant
  nature of the continuum limit spectrum  is to perform yet another
field redefinition: \beq \label{fermionmixing} \psi_{k  \alpha}
&=& {1\over \sqrt{2}}  \left( \sigma^3_{\alpha \dot\alpha}
\bar\chi_k^{\dot\alpha} + i \eta_{k  \alpha} \right) , \qquad
\bar\psi_k^{\dot\alpha} = {1\over \sqrt{2}} \left( - \bar\sigma^{3
\dot\alpha \alpha} \chi_{k \alpha} - i \bar\eta _k^{\dot\alpha}
\right) , \nnn \lambda_{k  \alpha} &=& {1\over \sqrt{2}}  \left(
\chi_{k  \alpha} + i \sigma^3_{\alpha \dot\alpha}
\bar\eta_k^{\dot\alpha}\right), \qquad
  \bar\lambda_{k}^{\dot\alpha} =  {1\over \sqrt{2}}  \left
( \bar\chi_k^{\dot\alpha} + i \bar\sigma^{3 \dot\alpha \alpha}
\eta_{k \alpha} \right). \eeq and write the final result for the
fermion bilinear lagrangian (\ref{fermion1}) in Fourier space as:
\beq \label{fermion6}
  L_{ferm.} &=& {1 \over g_3^2} \sum\limits_{k
= -{N-1\over 2}}^{N-1\over 2} \left[ {\over} - i \( \bar\eta_{k}^a
\bar\sigma^m \partial_m \eta_{k}^a + \bar\chi_{-k}^a \bar\sigma^m
\partial_m \chi_{-k}^a \) + |v| \sin {\pi k \over N} \left(
\bar\eta_{k}^a \bar\sigma^3 \eta_{k}^a
+ \bar\chi_{-k}^a \bar\sigma^3 \chi_{-k}^a \right)  \right]  \nonumber \\
&+& {1 \over \sqrt{N} g_3^2} ~\sum\limits_{k, k^\prime = -{N-1\over
2}}^{N-1\over 2} \left[ -{i\over 2}~ f^{abc} B^{c, k,
k^\prime}_\mu \( \bar\eta_k^a \bar\sigma^\mu \eta_{k^\prime}^b +
\bar\chi_{-k}^a \bar\sigma^\mu \chi_{-k^\prime}^b \) \right. \ddd
\left. + {1\over 2} ~f^{abc}~ \left( \eta_k^a \chi_{k^\prime}^b
~\phi_{k, k^\prime}^{* c} + \bar\eta_k^a \bar\chi_{k^\prime}^b
~\phi_{k,k^\prime}^c \right) \right] ~. \eeq
In the continuum limit ($|v| \to \infty$) all modes except
those with $k \ll N$ decouple and
the lagrangian (\ref{fermion6}) becomes that of
the 4d $N=2$ supersymmetric Yang-Mills theory.
We have defined the
Fourier components (along the latticized direction) of the 4d gauge
field  $B_\mu$ and the 4d complex adjoint scalar $\phi$ as
follows: \beq \label{4dfields} B_m^{c, k, k^\prime} &=&
A^{c,k - k^\prime}_m~ \cos^2 \( {\pi (k - k^\prime) \over
2N} \) , \nnn B_3^{c, k, k^\prime} &=& -{i \over \sqrt{2}} \left(
\cos{\pi k \over N} ~q^{*, c}_{k^\prime - k} - \cos{\pi k^\prime
\over N} ~q^c_{k - k^\prime}\right) ~, \nnn \phi_{k, k^\prime}^c
&=&  A^{c,k-k^\prime}_3 \cos^2 \( {\pi (k - k^\prime) \over
2N} \) \nnn && \qquad - { i \over \sqrt{2}} \left(\cos{\pi k \over
N}~ q^{* c}_{k^\prime - k} + \cos{\pi k^\prime \over N} ~ q^c_{k -
k^\prime} \right)~. \eeq
Note that the third component of the 4d gauge field, $B_3$, is the
imaginary part of the  link fields' fluctuations, $q_i$ of
(\ref{links}), while the scalar in the 3d vector multiplet, $A_3$,
combines with the real part of the link fields   to form the complex 4d
adjoint scalar $\phi$.    Some further comments are in order:
\begin{enumerate}
\item
The fields $\eta_k$, $\eta_{-k}$ and $\chi_{-k}, \chi_{k}$ in
(\ref{fermion6})  are the left and right moving components of the
two 4d adjoint fermions in the continuum limit, respectively. The
free part of the fermion action respects an $U(2)^N$ symmetry,
with $(\eta_k^a,\chi_{-k}^a)^T$ transforming as a doublet of
$SU(2)_k$  with unit charge   under the corresponding $U(1)_k$.
\item
The fermion bilinear  lagrangian $L_{ferm.}$ preserves, even for
finite $N$,
  a diagonal $SU(2)  \times U(1)  \subset U(2)^N$ symmetry,
with $k$-independent action on the fermions as described in the
previous  paragraph.
Note that, even though the current associated with this diagonal
$U(1)$  has the correct low-energy limit to represent (the 3d part of) a
4d
$U(1)_R$ current, it does not obey the correct anomalous Ward
identity. As will be shown below, there exists a current with the
same $R$ charges for the light states, but with different
$R$-charges for the heavy states, which gives the correct anomaly.
\item
We should stress that  the above diagonal $SU(2) \times U(1)$
symmetry of the fermion bilinear terms  is not a symmetry
of the lagrangian and is broken by the bosonic terms---it is
evident from the definition (\ref{4dfields}) of the 4d scalar adjoint
$\phi$ that one can  not  even consistently define an $U(1)$
action on $\phi$ for finite $N$ (so that  the kinetic terms of
$A_3$ and $q$ are invariant). We thus conclude that the classical
$SU(2) \times U(1)$ R-symmetry of the 4d $N=2$ theory is only
exact in the continuum limit.
\item
As usual in deconstruction, the mass spectrum can be read off
from eqn.~(\ref{fermion6}). For $k \ll N$, it approximates the
Kaluza-Klein spectrum of a   four-dimensional theory compactified
on a circle of radius: \beq \label{radius} R = {N \over \pi |v|}~.
\eeq The value of the lattice spacing is $a = 2/|v|$.
\end{enumerate}
The above analysis applies also to models of lattice supersymmetry
with two or more latticized dimensions, but the classical action
and field redefinitions turn out to be a lot more complicated and
there is no point of giving them here---as we will see (see section
3.4), the
D-brane picture makes the finite-$N$ violation of the chiral
symmetries explicit.

\subsection{Equivalence to Wilson Fermion Action and
the   Anomaly Calculation}

We now go back to the $x$-space lattice lagrangian
(\ref{fermion1}). It has already been noted that deconstruction of
5d theories gives rise to Wilson terms for the fermions\footnote{One of
us (E.P.) thanks Chris Hill for
discussions.}
\cite{Hill:2002me}; this, in fact,  is a generic feature of
deconstruction. Here, we show this using the deconstructed theory
action  obtained  in section 2.1. More generally, in section 3.4, we
will see that the Wilson term has a simple geometric interpretation in
the brane realization of deconstruction.

Let us
compare our lattice fermion action (\ref{fermion1}) to the one
used in the work on anomalies with Wilson fermions  of Karsten and
Smit (KS) \cite{KS81}. We start with the usual Wilson fermion
action on a 4d lattice (see e.g. \cite{KS81}). We convert to a
2-component notation, where the Dirac spinor $\psi$ of KS
is\footnote{$\psi$, $\eta$ and $\chi$ here should not be confused
with the spinors of the previous subsection.} \beq \psi =
{\chi_\alpha \choose \etab^\adot} ~, \label{dfde} \eeq set   the
bare mass to zero, take  the continuum limit in $x^0, x^1$ and $x^2$,
and  obtain the Wilson fermion lagrangian with one latticized
dimension: \beqa L_{ferm.}^{KS} &=& \sum_j \left\{ {\over} -i \[
\etab_j \sbar^m \p_m \eta_j + \chib_j \sbar^m \p_m \chi_j \]
\right. \ddd -{i \over 2a} \[ \etab_j \sbar^3 (\eta_{j+1} -
\eta_{j-1}) + \chib_j \sbar^3 (\chi_{j+1} - \chi_{j-1}) \] \ddd
\left. + {r \over 2a} \[ \eta_j (\chi_{j+1} + \chi_{j-1} - 2
\chi_j ) + \etab_j (\chib_{j+1} + \chib_{j-1} - 2 \chib_j ) \]
\right\} . \label{laff} \eeqa We compare this to the free part of
\myref{fermion1}: \beqa L^{free}_{ferm.} &=& {1 \over g_3^2}
\sum_j \left\{ -i \[ {\vs \over 2} \psib_j \sbar^m \p_m \psi_j +
\lb_j \sbar^m \p_m \lambda_j \] \right. \ddd \left. - {i \vs \over 2
\sqtw} \[ \psi_j (\lambda_{j+1} - \lambda_j) - \psib_j (\lb_{j+1} - \lb_j )
\] \right\} . \label{ffmo} \eeqa
Recall that the lattice spacing is
$a=2/\va$. The field redefinitions  relating the deconstructed theory
lagrangian  (\ref{ffmo}) to
the Wilson fermion lagrangian (\ref{laff})
are:\footnote{Note that these are related to \myref{fermionmixing}
by $k \to j$ and $\eta \to -\eta$; the sign is important in the
Wilson mass term.} \beqa \label{redef3} {\va \over g_3 \sqtw}
\psi_{j\alpha} &=& {1 \over \sqtw} (\saat \chib_j^\adot - i
\eta_{j\alpha} ) ,
\\
- {1 \over g_3} \lambda_{j\alpha} &=& {1 \over \sqtw} (\chi_{j\alpha} -
i \saat \etab_j^\adot ) .\nonumber \eeqa The matching of the two
lagrangians shows that $r=1$ in the deconstructed theory. This
value of the Wilson term is imposed by the explicit supersymmetry
and gauge invariance of the deconstructed theory.

In the four-component basis of \myref{dfde}, the axial $U(1)$
current is associated with $\psi \to e^{-i \vp \gamma_5} \psi$
with $\gamma_5 = \diag(-1,1)$. Making this local we have in the
2-component notation: \beq \eta_j \to e^{i \vp_j} \eta_j, \qquad
\chi_j \to e^{i \vp_j} \chi_j . \eeq This axial  transformation in
our deconstructed theory is the  anomalous $U(1)_R$ transformation
of the continuum 4d theory, combined with a diagonal generator of
the anomaly free $SU(2)_R$.

The variation of the lattice action
   \myref{laff}, or, equivalently, (\ref{fermion1}) written in terms of
the new variables (\ref{redef3}),  under the axial $U(1)$
is:\footnote{Of course,
$\delta L^{free}_{ferm.} / \delta \vp_j$  of
eqn.~(\ref{fermion1})  is obtained from this by taking $r=1$.}
\beqa {\delta L^{KS}_{ferm.} \over \delta \vp_j} &=& - \p_m \[
\etab_j \sbar^m \eta_j + \chib_j \sbar^m \chi_j \] \ddd -{1 \over
2a} \[ \( \etab_j \sbar^3 \eta_{j+1} + \chib_j \sbar^3 \chi_{j+1} +
\hc \) - \( \etab_{j-1} \sbar^3 \eta_j + \chib_{j-1} \sbar^3 \chi_j
+ \hc \) \] \ddd + {r \over 2a} \[i\eta_j (\chi_{j+1} + \chi_{j-1}
- 2 \chi_j ) + i\chi_j (\eta_{j+1} + \eta_{j-1} - 2 \eta_j ) + \hc
\] . \eeqa In the continuum limit, the second term in brackets
becomes: \beq - \p_3 \[ \etab \sbar^3 \eta + \chib \sbar^3 \chi \] .
\eeq Thus the anomalous Ward identity is given by: \beqa
\lefteqn{\bigvev{\p_\mu \[ \etab \sbar^\mu \eta + \chib \sbar^\mu
\chi \]}_{x^3_{(j)}} } && \ddd = \lim_{a \to 0} {r \over 2a}
\bigvev{
\[i \eta_j (\chi_{j+1} + \chi_{j-1} - 2 \chi_j )
+ i \chi_j (\eta_{j+1} + \eta_{j-1} - 2 \eta_j ) + \hc \] } .
\eeqa The right-hand side is nothing but the ``axial symmetry
breaker $D^A_0$'' of KS.\footnote{Up to an overall normalization.
Also note that in the notation of KS, $\lambda_0 = 1$ corresponds
to the flavor-diagonal current.} To obtain this result, note that
$M=r/a$ (in the notation of KS, eqn.~(5.7)  of \cite{KS81}) since we
only have a Wilson term in one
dimension and the bare mass is zero.  In addition, we make use of the
identity $- \psib_j \gamma_5 \psi_\jp = \eta_j \chi_\jp -
\chib_j \etab_\jp$ for the Dirac spinor \myref{dfde} taken at
different points $j, \jp$ along the latticized dimension.

Having thus established the equivalence of the deconstructed
theory action to
  the    Wilson fermion action and having matched  the corresponding
anomalous Ward identities, we are assured that the
remainder of the weak coupling axial anomaly calculation goes
through according to KS. The latticization of one
dimension regulates the linear divergence usually associated with
the weak coupling anomaly; furthermore the regulator respects the
vector current Ward identity and Bose symmetry; hence we are
assured that the axial anomaly emerges just as if we had
latticized all 4d, as KS have done.

One difference with the work of KS is worth pointing out. In a
nonsupersymmetric lattice theory,  one has the
  freedom to vary the Wilson term coefficient $r$ at will.  For $r = 0$,
the doublers have a continuum 4d interpretation as massless fermions of
opposite axial charge which cancel  the anomaly of the  physical
fermions. For $0 < r < 1/\sqrt{2}$,  the doublers still have a 4d
continuum interpretation as fermions of opposite axial charge, but
with chiral symmetry violating masses. This follows from the Wilson
fermion dispersion relation, which can be read off eqn.~(\ref{laff}),
or, more easily, from the geometric D-brane picture, see
eqns.~(\ref{geometric}), (\ref{wilson3}):
\beq
\label{wilson1}
a^2 \omega^2(k)  =   \sin^2 {2 \pi k\over N} + 4 r^2 ~   \sin^4 {\pi
k\over N} ~.
\eeq
(Here $\omega$ is the energy and for simplicity we have set the continuum
momenta $p^1,p^2$ to zero.)
Thus (for $0 < r < 1/\sqrt{2}$) the  effect of the doublers can be
``physically" described as that of Pauli-Villars regulators, whose
axial symmetry violating mass terms give rise to the anomaly. In
our case, $r=1$, it is easy to see that there is no continuum
interpretation of the would-be ``doublers" as 4d massive Dirac fermions;
nevertheless, as KS showed, the calculation of  the anomaly is
valid for all values of $r$.

\mysection{Fractional Branes and the Chiral Anomaly via Classical
Supergravity}
  In this section, we show that the lattice
perturbation theory result   for the $U(1)_R$ anomaly can be
obtained via a classical supergravity calculation in the
fractional brane realization of the lattice supersymmetric models.
The brane construction is very general and underlies all proposed
realizations of lattice supersymmetry by deconstruction.

For completeness, we begin in section 3.1 by reviewing
  the fractional brane construction of the continuum limit 4d $N=2$ theory
  and describing the calculation of the continuum theory $U(1)_R$
  anomaly. This section also reviews well-known results about the
  closed string calculation of the running coupling in supersymmetric
gauge
  theories.

The fractional brane construction of the deconstructed $SU(k)^N$
theory is given in section 3.2. Then, in section 3.3, we study the
running of the gauge coupling in the deconstructed theory. We
compare the running coupling calculation (section 3.3.1) with the field
theory (section 3.3.2)
    and  study the continuum limit in the fractional
brane picture. The supergravity calculation of the anomaly is discussed
in section 3.3.3.
Finally, in section 3.4, we give a geometric picture of the R-symmetry
and the
symmetry-breaking Wilson term.

\subsection{Anomalies in Seiberg-Witten Theory via Fractional Branes}

First, let us derive the non-abelian anomaly and $\beta$-function
of the Seiberg-Witten  theory from an appropriate orbifold. This
will also be the desired result in the continuum limit of the
deconstruction setup.

In order to realize a four dimensional theory with eight
supercharges, we  take D3 branes stretched along $X^0,..,X^3$.
The D3 branes are   transverse to a $C^2/Z_r$ orbifold acting in the
standard way on the
coordinates $X^4,...,X^7$. Here $r \geq 3$, but otherwise
arbitrary.

This gives a quiver gauge theory \cite{Douglas:1996sw} on the
brane, whose matter content depends on the choice of
representation of the Chan-Paton (CP)  factors. The simplest such
choice, namely the one-dimensional trivial representation of $Z_r$,
gives a pure gauge theory.\footnote{Obtaining a theory with matter
of the quiver form is done simply by including more
representations of $Z_r$. This should not present any essential
new features for our purposes here.}  Note that the more common
choice---the regular representation---is $r$-dimensional,   allowing
for branes to move away from the fixed point. Here the brane is
stuck at the fixed point (i.e. its world-volume theory has no Higgs
branch), and is dubbed a
``fractional brane.'' Of course one can have any multiplicity for
these representations, giving non-abelian theories.

These fractional branes behave slightly differently from the usual
D3 branes \cite{fractional}. For example, when calculating their
kinetic term, it is a different modulus than the complexified
string coupling that controls their gauge coupling. This can be
demonstrated by an explicit worldsheet calculation, as sketched
below.

The closed string sector has $r-1$ twisted sectors, and each one
of them has in the spectrum a  massless complex scalar
(propagating in the 6 dimensions transverse to the orbifold). We
denote  these moduli by $\tau_i$, where for a general orbifold $i$
runs over all the nontrivial elements of the discrete group. Here
$i=0,...,r-1$, where  we denote by $\tau_0$  the usual
(untwisted) complexified string coupling.

Generally, to specify a particular quiver theory one chooses a
representation of the CP factors of some dimension, say
$m$. Here $m=1$, while for the regular representation $m=r$ (we discuss
here
the general
case, which will be useful later). This means  that every
element of the group is represented by a matrix $\gamma_i$ of
dimension $m$. The gauge fields carry a (diagonal) $m \times m$
matrix $A$. For non-abelian theories we simply take some
multiplicity of the above structure.

One can now proceed to calculate the gauge theory action from
various worldsheet couplings. The actual worldsheet amplitude will
be only a (non-zero) irrelevant factor. All that matters is
various selection rules one gets from the CP part of the
amplitudes.

For example,  in calculating kinetic  terms in the action, one has
to calculate the amplitude with two open string vertex operators
and one closed string modulus. The dependence on the CP
factors is encoded in the trace Tr$(A \gamma_i A)$. For example,
the usual case---with the regular (traceless) representation  for
$\gamma_i$---has $A$ proportional to the unit matrix, therefore
Tr$(A \gamma_i A)=0 $ unless $i=0$. One then deduces that $\tau_0$
is the gauge coupling on the brane.

For the fractional brane, on the other hand, the dimension $m=1$ and $\gamma_i =
\omega_r^i$, where
$\omega_r$ is any $r$-th root of unity.  If we choose for
simplicity this root of unity to be $1$, then all the generators
of $Z_r$ are simply represented by that number. With this choice,
our fractional brane couples identically to all
fields $\tau_i \; (i=0,\ldots,r-1)$.  This is
more conveniently summarized if we perform a discrete Fourier
transform on the closed string fields: \beq \label{coupling1}
\tau_m = \sum_{i=0}^{r-1} ~\omega_r^{im} ~\tau_i~. \eeq
   We see
therefore that our fractional brane couples to the field
$\tau_{m=0}$ only. The other possible choices of $r$-th root of
unity lead to fractional branes, each of  which couples to a
specific Fourier mode of the closed string fields.

Having identified the holomorphic gauge coupling, we are now ready
to discuss quantum effects. Suppose one wants to calculate the
running coupling of a non-abelian gauge theory $SU(k)$ at some
scale $|A|$. This can be achieved, for example, by going to a particular
point of
its Coulomb branch, where a VEV of magnitude $|A|$ breaks the
gauge symmetry to $SU(k-1) \times U(1)$. The low-energy effective
action (valid at scales below $|A|$) then contains a kinetic term
for the abelian field:
   \beq \int d^2\theta \, \tau(A) \,W^\alpha W_\alpha, \eeq
and the value of the  $U(1)$ (frozen) gauge coupling  is equal to the
$SU(k)$ (running) gauge coupling, evaluated at
the scale of the breaking $|A|$.

In the field theory, the abelian  gauge coupling at $A$ (and below)
  is obtained by integrating out the massive W-boson supermultiplets
which are charged
under the $U(1)$.  In a brane realization  the $SU(k)$
symmetry lives on a set of $k$ D-branes, which we presently separate to
one group of $k-1$ branes, which we call the ``source branes,'' and
   a single brane, called the
``probe brane.'' The massive W-bosons are then strings stretched
between the source and the probe ($s-p$ strings). We emphasize that
the distinction between the source and the probe is a matter of
convenience, and applies  even in the $k=2$ case.

  At one loop, the  result  can be extracted from an annulus diagram
with two massless open-string vertex operators, corresponding to
$U(1)$ gauge bosons, inserted on the boundary.
  In order to isolate the contributions of the $s-p$
strings alone, the boundary conditions on both boundaries are chosen
differently:  one of them (the one with the gauge vertex operator
insertions) in correspondence to the probe brane, and the other
in correspondence to the source branes. The mass of the W-bosons is
proportional to the separation between the branes, so in order to
decouple the open-string oscillator modes, one works in the limit
of sub-stringy separations.

The supergravity analysis relies on the old observation that the
annulus diagram can be calculated in the closed-string channel as
well. The  limit appropriate for a supergravity  treatment is that
of large separations compared to the string length, but
supersymmetry \cite{Douglas:1996du}
aids in making the extrapolation to the sub-stringy, field theory,
regime.

In the closed-string channel, the annulus factorizes into a sum of
all possible intermediate closed string fields. In the
supergravity limit only the massless fields are relevant. Each
contribution is then a product of three factors:
\begin{enumerate}
\item
  The tadpole: a disc diagram with one closed-string insertion, and
boundary
  conditions corresponding to the
  source branes.  In the supergravity  action this is
summarized by a source term for the closed-string field. From the
above discussion we see that only the field $\tau_{m=0}$  has a
tadpole. The tadpole in our case is proportional to $k$, the total
number of  branes.\footnote{And not to $k-1$. This comes  from the
contribution of CP factors  to the factorized closed-channel
amplitude: the source's coupling to the tadpole has a factor of
Tr $1 = k$.} We choose to normalize the closed-string fields such
that the tadpole is $\frac{b_0}{ 2 \pi} = \frac{k}{\pi}$, where
$b_0 =2k$ is    the one-loop $\beta$ function coefficient of
$SU(k)$.
  \item
  Closed-string propagator:  since the twisted sector fields are
  allowed to vary only in two directions, this gives a logarithmic
  dependence on the coordinates.\footnote{In our conventions for the
normalization
  of the closed-string field kinetic terms,
   the (2 dimensional) propagator satisfies
  $ \nabla^2 G_2(x,y) = \,  2\pi \delta(x) \delta(y)$.} Using complex
notation for the $X^8, X^9$ plane,
$z \equiv X^8 + i X^9$,
one gets:
    \beq
\label{source}\Phi(z) =  c + \frac{b_0}{ 2 \pi}\, \log (|z|)  ~,\eeq
    where $c$ is an integration constant. $\Phi(z)$ is a NS  sector
    scalar, which we relate to the coupling constant below.
\item The response: a disc diagram with two open-string vertex
operators, and one closed-string vertex operator. This summarizes
the coupling of the probe to the closed-string field. The
interaction of the probe occurs at the location $z=A$, and its
strength can be chosen to be $\frac{1}{4\pi}$ (by normalizing the
open string gauge fields appropriately), therefore the coupling on
the probe is:  \beq \label{running} \frac{1}{g_{eff}^2}
= \frac{c}{4 \pi}+ \frac{b_0}{8 \pi^2}\log (|A|)~.\eeq
This indicates that the integration constant $c$ should be related to the
bare coupling by $\frac{c}{4\pi} = \frac{1}{g_0^2}$.
\end{enumerate}

We can use  holomorphy to deduce the imaginary part of the
coupling (\ref{running}). Defining as usual $\tau(A) =
\frac{\Theta(A)}{2\pi} +
i \frac{8 \pi^2}{g^2(A)}$, one gets, \beq \tau(A) = \tau_0(A) + i
\, b_0 \log A . \eeq
  In the fractional brane realization of the $N=2$
theory, the anomalous $R$-symmetry is the $SO(2)$ rotation in
$X^8,X^9$ directions, i.e. $z \to e^{i \alpha} z$ implies
$A \rightarrow e^{i \alpha} A$.  (The
$SU(2)_R$ comes from action on the orbifold directions, or more
generally on the complex structures of $K3$.) Therefore the result
(\ref{running}) encodes both the running coupling and the
R-symmetry
   violation.

\subsection{ The Deconstruction Setup with Fractional Branes}

We now consider   the deconstruction of the same theory.
Generally, there are two ways  to view deconstruction as a string
theory setup:
\begin{enumerate}
\item
As an orbifold of a Hanany-Witten setup \cite{Lykken:1997gy}, where
D-branes  are stretched between NS
fivebranes. In the continuum limit (moving asymptotically on the
Higgs branch) the D-branes combine and move away from the orbifold in
one direction along the NS
fivebranes. We will not need this setup below, so
we will not provide any details here.
\item
  As a setup of fractional branes on an orbifold. This is the description
we  adopt as it allows
  for a
perturbative string theory treatment \cite{Leigh:1998hj}.
\end{enumerate}
In our case---the deconstruction of 4d Seiberg-Witten theory---we
start with   D2 branes\footnote{The spatial lattice constructions of
\cite{Kaplan:2002wv}
can be obtained by   similar (fractional) brane constructions: the 3d
$N=4$ theory is obtained
upon replacing our fractional D2 branes with fractional D0 branes, and
the $C^2/Z_r \times Z_N$ orbifold by a
$C^3/Z_r \times Z_{N_1} \times Z_{N_2}$ orbifold;  the 4d $N=4$ theory,
on the other hand,  involves only regular D0 branes on a  $C^3/ Z_{N_1}
\times Z_{N_2} \times Z_{N_3}$ orbifold.}
stretched along the $X^0,X^1,X^2$
directions. The orbifold group is chosen to be $Z_r \times Z_N$
(we take $r,N$ to be relatively prime), which acts as follows:
\begin{itemize}
\item
The generator of $Z_N$ acts on the directions $X^3+ i X^8, X^4 + i
X^5$ in the standard way.  In the deconstruction setup one needs to
move into the Higgs branch, so our branes will need to be   in the
regular representation of $Z_N$.
\item
The generator of $Z_r$ acts   on the directions $X^4+ i X^5, X^6 + i
X^7$. The CP assignments for this generator are as before
(section 3.1) only taken with multiplicity of $k N$, to account
for all images with respect to the $Z_N$ orbifold action, and to
generate a nonabelian $SU(k)$ gauge group in four dimensions.
\end{itemize}
Thus, our CP matrices are $kN$ dimensional. Let us work out the CP
factors corresponding to the various bosonic fields. Denote the
action of the generator $Z_N$ on the CP factors by $\gamma_N$ and
take the regular representation, tr $\gamma_N = 0$: \beq
\label{regulargamma} \gamma_N = {\rm diag} ( \omega_N, \omega_N^2,
\ldots , \omega_N^N)~, \eeq where every element is multiplied by a
$k \times k$ unit matrix. The gauge field's ($A_{m}$, $m = 0,
1,2$) CP factors are (generally complex)  $k N \times k N$
matrices, which we denote by ${\bf A}$. They obey: \beq
\label{gaugeorbifold} {\bf A} = \gamma_N {\bf A} \gamma_N^{-1}~,
\eeq which means that the states that are not projected out by the
$Z_N$ orbifold
  have block diagonal CP factors:
\beq \label{gaugeNorbifold}
{\bf A} = \left( \begin{array}{ccccc} \lambda_1 & 0 & \ldots & 0 & 0 \\
0 & \lambda_2 & \ldots & 0 & 0 \\
0 & 0 & ...\lambda_i ... &0&0 \\
0&0&\ldots & \lambda_{N-1} & 0 \\
0&0&\ldots &0 & \lambda_N \\ \end{array} \right)~, \eeq where the
$\lambda_i$ are the $k \times k$ CP matrices of the unbroken
$U(k)^N$ gauge groups.

Similarly, we take as the generator of the $Z_r$ action $\gamma_r
= {\rm diag}(\omega_r, \ldots, \omega_r)$, with $\omega_r$ the
one-dimensional representation of $Z_r$. It is evident that all
gauge boson states with CP factors (\ref{gaugeNorbifold}) survive
the $Z_r$ orbifold projection.

We now move on to describing the scalar fields. One  scalar field
that survives the combined $Z_N \times Z_r$ action  corresponds to
motion of the D2 branes in $X^9$ (the scalar of the 3d $N=2$ vector
multiplet). This  scalar is described by the
  same CP factor as the gauge field (\ref{gaugeNorbifold}).

  The other set of surviving scalars correspond to moving in the  $X^3 +
i X^8$ directions, and make up
  the link fields (bifundamentals) of the deconstruction setup.
  The link fields CP factors ${\bf \Phi}$ obey:
\beq \label{linkCP1} {\bf \Phi} = \omega_N \gamma_N ~{\bf \Phi}~
\gamma_N^{-1}~, \eeq which is solved by: \beq \label{linkCP2}
{\bf \Phi} = \left( \begin{array}{cccccc} 0 & \phi_{1,2} & 0 & \ldots &
0 & 0 \\
0 & 0 & \phi_{2,3} &\ldots & 0 & 0 \\
0 & 0 & 0 &...\phi_{i,i+1}...&0 & 0\\
0 & 0 & 0 & \ldots &0 & \phi_{N-1,N} \\
\phi_{N,1}&0&0&\ldots &0 &0\\ \end{array} \right)~, \eeq where
$\phi_{i,i+1}$ are the $k\times k$ CP matrices of the $N$ link
fields. In fact, the CP factors of the link fields above
correspond to our   fields $\phi_{i,i+1} \rightarrow Q_i$ of
eqn.~(\ref{kahler}) in the
coordinate representation.

The moduli space of the $k N$ D2 branes on the orbifold is
three-dimensional---branes are allowed to move in groups of $k$,
in a $Z_N$ symmetric way in $X^3 + i X^8$ (Higgs branch), as well as
in $X^9$ (Coulomb branch).  The deconstructed theory is obtained
upon moving them a distance $v$ from the origin in $X^3 + i X^8$,
i.e. along the Higgs branch, as explicitly described in  section
2.1.

To proceed with the analysis of gauge coupling running and the
anomaly, we need the couplings of the world-volume open string
massless  modes
  to the closed string fields.  This is a particular example of
compactification of type IIA
  string theory on a Calabi-Yau manifold, and the closed string spectrum
is well-known (see e.g.
\cite{Douglas:1996sw});  we review
  the orbifold analysis for the sake of completeness.

   A priori there are $Nr$ twisted sectors of closed strings to
consider. However, to be sourced by our D-brane setup, the closed
string fields need to be a particular combination of the twisted
sectors with respect to the $Z_r$ orbifold, as described in the
previous section.

  Therefore, for our purposes there are only $N$
non-trivial sectors to consider.
  These are $N-1$ twisted sectors, and
one untwisted sector, with respect to the $Z_N$ action.  To see
the matter content in these sectors it is useful to  ignore the
$Z_N$ orbifold at first. In this case the twisted sector fields
fill out six dimensional $(1,1)$ vector multiplets.\footnote{The
untwisted sector contains also a 6 dimensional gravity multiplet,
which decouples from our analysis.} The bosonic fields in each one of
those
multiplets consist of a real vector (from the R-R sector) and 4
scalars (from the NS-NS sector).

Subsequently imposing the $Z_N$ projection yields a
3-dimensional $N=4$ vector multiplet for each of the $N$ sectors
relevant for us.\footnote{The untwisted sector with respect to
$Z_N$ contains an additional hypermultiplet, which is irrelevant for our
analysis.} The bosonic closed string fields surviving the
complete orbifold projection are then  complex scalar fields
$\Psi_i$ (from the NS-NS sector), a  vector field $V_i$ (from the
R-R sector) and a real 0-form $S_i$ (from the R-R sector), where
$i = 0,...,N-1$. As usual, the $i=0$ case refers to the untwisted
sector  with respect to $Z_N$, and as mentioned above all the
fields are assumed to be the correct linear combination of twisted
sectors with respect to the $Z_r$ action (the $m=0$ Fourier mode,
in the notations of the previous section, eqn.~(\ref{coupling})).

The charges of the D-branes with respect to these fields can be
calculated  by a disc diagram with no external states and a closed
string twisted state vertex operator insertion. Since we are
interested in the configuration of separated fractional branes, we
calculate the charges of each fractional brane separately.

The only new element here is the factor that comes from the CP
matrices of the $Z_N$ action. If we are interested in the $l$-th
fractional brane, then the generator of the $Z_N$ action is
represented by the CP factor $w_N^l \, 1_{k \times k}$. Therefore
the coupling to any of the  $q$-th twisted sector (with respect to
$Z_N$) closed string fields is proportional\footnote{Note that we
choose the trace over the CP matrices to be normalized such that
Tr$(1_{N \times N}) =1$. In particular, in these conventions the
charge of a brane in the regular representation (which has all $N$
images) is chosen to be $1$.}
  to $\frac{k}{N} ~w_N^{ql}$.

  The four dimensional couplings, in the continuum limit, are T-dual to
  the closed string fields in our brane setup. It is clear then
  that we are interested only in the untwisted sector fields $q=0$ (with
  respect to $Z_N$).\footnote{These fields are the parameters of the 3
dimensional action.}
   We denote these fields by $\Psi$ (a complex scalar field),
  $V$ (a vector field), and $S$ (a 0-form).
  We choose the normalizations of the closed string fields such that the
coupling of the
  individual fractional
  branes to these fields is $\frac{b_0}{N}$, with no phase
  factors. This follows from the standard    relation between
  different brane tensions
  ($T_2 = 2\pi T_3$, for $\alpha'=1$).

\subsection{The  Running Coupling and the Anomaly}

\subsubsection{ Supergravity Calculation}

Let us elaborate on the equations of motion for the closed string
fields  sourced by the D2 brane configuration (shown in Figure 1).
The sources for the
tadpole are
   localized in the $X^3, X^8, X^9$ space at the locations:
   \begin{eqnarray}
\label{sources}
   X^3_k &=& v \sin {2\pi k \over N} ~,\nonumber \\
   X^8_k &=& v \cos {2\pi k\over N} ~,\\
X^9_k &=&0 \nonumber~,
   \end{eqnarray}
i.e. the branes are distributed with equal spacing  on a circle of
circumference   $2 \pi v$ in the $X^3$--$X^8$ plane.
\begin{figure}[h]
\begin{center}
\includegraphics[height=3.0in,width=3.0in]{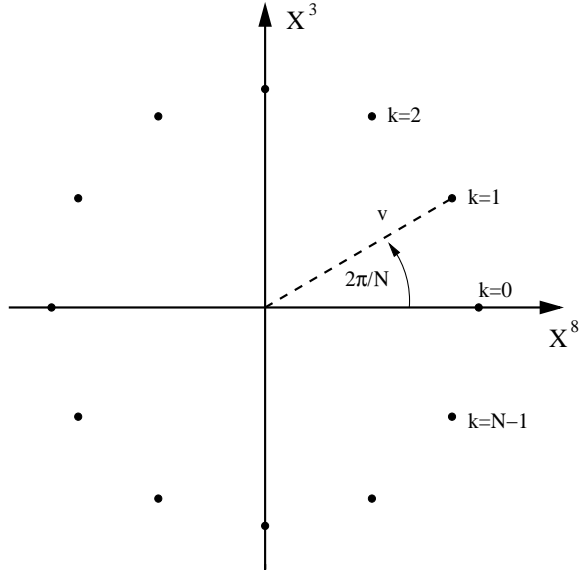}
\end{center}
{\caption{ \small D-brane sources of the $Z_r$-twisted fields  $\Psi$,
(\ref{tadpole0}), and $U$, (\ref{vector}), in a $Z_N$-symmetric
distribution along a circle of radius $v$ in the $X^3$-$X^8$ plane, at
$X^9=0$.
The Coulomb branch direction $X^9$ is perpendicular to the drawing.
  The branes are labeled by
values of $k$, as defined in (\ref{sources}).}}
\label{f1}
\end{figure}
  Each one of the source terms is of strength $\frac{b_0}{ N}$ in the
  normalization of the last section.
The NS-NS  sector scalar  field $\Psi$  couples to the branes as
electric sources,\footnote{We choose to normalize coordinates such
that the 3d propagator  satisfies $\nabla^2 G_3 = \,\delta(x)
\delta(y) \delta(z)$; therefore we have in comparison to the
previous section $\int\limits_{-R}^{R} dz\, G_3(x,y,z) \simeq 2\pi
\, G_2({x\over R},{y\over R})$ for $x,y \ll R$.}   i.e.: \beq
\label{tadpole0} \nabla^2 \Psi =  \frac{b_0 }{N} ~\sum_{k=0}^{N-1}~
\delta^{(3)} (X-X^k)~, \eeq
which yields for the value of $\Psi$ at arbitrary $X^{3}, X^8, X^9$:
  \beq
\label{tadpole1}
  \Psi(X^3, X^8, X^9) =   C - \frac{b_0 }{4 \pi N} ~\sum_{k=0}^{N-1}~ {1
\over \sqrt
{ (X^3 - X^3_k)^2 + (X^8 - X^8_k)^2 + (X^9)^2 }}~.
  \eeq
  where $ C$ is a constant.

In order to see a 4d behavior, we need to  focus to a neighborhood
close enough to the circle of D-branes, such that in the large-$N$,
large-$v$ limit, a nearby section of
the circle approximates a straight
line. To that effect, we place the probe brane at a position close to
the sources. Since the probe brane is also in the regular
representation with respect to $Z_N$, this will include all images
of that brane

The continuum limit of the deconstructed theory is best exhibited
upon introducing appropriate, see Figure 2, coordinates\footnote{See also
discussion about coordinate choice in section 3.4} (everything
here and below is made dimensionless in terms of the string length
$l_s$) $\rho, \eta, \phi$ defined as: \beq \label{coordinates}
X^3 &=&  (v  + \rho) \sin \phi \nonumber ~,\\
X^8 &=&  (v  + \rho) \cos \phi ~, \\
X^9 &=& \eta \nonumber~. \eeq The probe brane
is placed\footnote{This choice of $\phi$-coordinate  for the position of
the probe
brane indicates that we do not turn on a Wilson line for the 4d gauge
field even at finite $R$,
since, in any case, it would vanish  in the continuum $R\rightarrow
\infty$ limit. } at
constant $ \rho, \eta$, and at $\phi = \frac{2\pi k'} {N} $, with
$k^\prime = 0,\ldots,N-1$ counting
  its images.
\begin{figure}[h]
\begin{center}
\includegraphics[height=3in,width=3.5in]{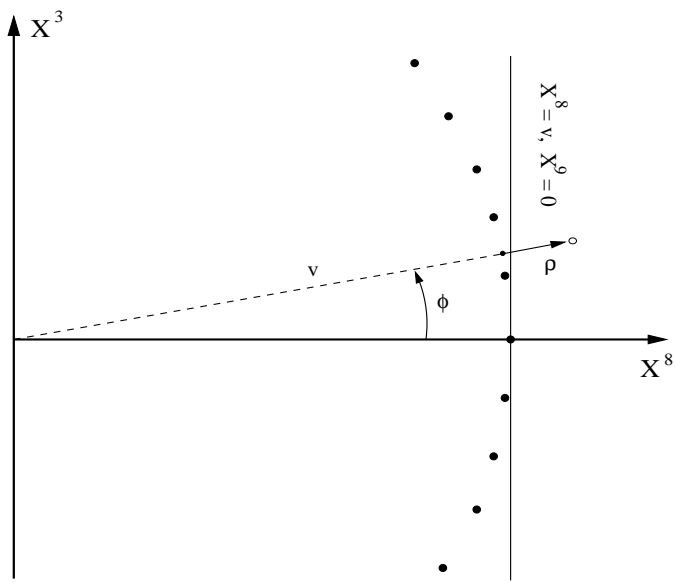}
\end{center}
\caption{ \small The location of a probe brane (at  general $\phi$),
shown here by an open circle, in relation to the arc of source branes
in the neighborhood of $X^8=v,~X^3=X^9=0$.  According to
(\ref{coordinates}), $\rho$ measures the radial
displacement of the probe brane from this arc (of
radius $v$), while $\phi$ gives the angular displacement
from the $X^8$ axis.  The $X^9$ displacement $\eta$
of the probe brane is not shown; the
figure should be thought of
as a projection onto the $X^9=0$ plane. The vertical line passing
through $X_8 = v$ is the R-symmetry axis of section 3.4.}
\label{f2}
\end{figure}

In contrast to the previous section,
the normalization of 3-dimensional gauge fields is
  chosen such that the probe couples to $\Psi$ with strength
  $\frac{1}{2}$. In particular, with these conventions the 3
  dimensional actions does not have an overall $\frac{1}{N}$
  factor.

In the absence of the source, the gauge coupling on the probe
brane is determined by the constant of integration, $C$ of
eqn.~(\ref{tadpole1}). With the source, one finds for the gauge
coupling on the probe:
   \beqa \label{tadpole2}
\frac{1}{g^2_{probe}} &=& \frac{1}{2} \sum_{k'=1}^N \Psi(\rho,
\eta, \phi= \frac{2\pi k'}{N})  \nonumber \\ &=& \frac{NC}{2} -
{b_0 \over 16 \pi   v}~ \sum\limits_{(k-k') = -{N-1 \over 2}}^{N-1
\over 2} ~ \left[ \left(1 + {\rho \over v}\right) \sin^2 \left({
\pi (k-k') \over N} \right) + {\rho^2 +
\eta^2 \over 4 v^2} \right]^{-{1\over 2}}~ \\
&\simeq& \frac{NC}{2} -   {b_0 N\over  16 \pi^2 v }  ~
\int\limits_{-{\pi\over 2}}^{\pi \over 2} {d x \over \sqrt{ \sin^2
x + {\rho^2 + \eta^2 \over 4 v^2}}} \simeq  \frac{NC}{2}+ { b_0 N
\over 8 \pi^2
   v} ~ \log {|\rho
+ i \eta| \over 2   v } \nonumber ~, \eeqa where we took    $1 +
\rho/v  \simeq 1$, and replaced the sum\footnote{See \cite{Sfetsos:1998xd}.}
by an integral in the
large-$N$ limit. Note that the last equality in (\ref{tadpole2})
is only valid for $|\rho + i \eta| \ll v$; the tadpole falls off
as an
  inverse power  of $|\rho + i \eta| $ at large values of the argument.

In the next section, we will see that the probe-brane coupling,
eqn.~(\ref{tadpole2}), reproduces the running coupling of the continuum
theory and will relate  the constant $C$ to the ``bare" 4d coupling.
In order to continue with the interpretation of eqn.~(\ref{tadpole2}),
we thus remind ourselves of the deconstructed field theory calculation
of the running coupling.

\subsubsection{Field Theory Calculation   and the Continuum Limit}

We want to relate the tadpole $\Psi(\rho, \eta, \phi$)
of eqn.~(\ref{tadpole2}) to the running of the gauge
coupling in the field theory, evaluated at the scale $|A| = |\rho +
i \eta|$. The gauge coupling on a  probe brane placed at $\rho +
i\eta$, away from the circle, will depend on the background
$\Psi$, as described in  section 3.2 for the continuum
limit theory. For simplicity we will only focus on the
$SU(2)^N$ case.

The probe brane gauge coupling is the coupling of the unbroken
$U(1)$ on the Coulomb branch of the diagonal $SU(2) \subset
SU(2)^N$ 3d gauge group---moving a brane away from the circle of
radius $v$ in the $X^3, X^8$ plane is equivalent to going on the
Coulomb branch. The running of the coupling is due to the heavy
gauge bosons and superpartners which acquire masses on the Coulomb
branch. The holomorphic gauge coupling ``runs" only at one
loop---more precisely, upon integrating out a   (3d $N=4$) vector
multiplet of mass $m$, the 3d diagonal gauge
coupling $g_{3,D}$ receives a threshold correction: \beq
\label{3dmatching} {1 \over g_{3,D}^2}_{low} = {{1 \over
g_{3,D}^2}_{high}} - {1 \over 2  \pi m}~, \eeq where the
coefficient was calculated in \cite{Dorey:1998kq}; the minus sign
indicates that the 3d theory becomes more  strongly coupled in the
infrared.

To calculate the gauge coupling of the $U(1)$ at the scale $A$, we
have to sum the thresholds (\ref{3dmatching}) from all heavy
states charged under the $U(1)$, the $W^{\pm}$ bosons. We start
with the tree level value of the diagonal gauge coupling at the
  cutoff scale\footnote{Hereafter, our normalization of the vev
(\ref{links}) of $Q_i$ differs from that of section 2: in the brane
picture, it is conventional to have $2 \pi R = N/v$, as opposed to
(\ref{radius}).} $v$:
\beq {1\over g_{3,D}^2} = {N\over g_3^2}~,
\label{cutoffcoupling}
\eeq  as is usual in
deconstruction.  A short calculation gives for the $W^{\pm}_k$
boson masses along the Coulomb branch, for $|A| \ll v$: \beq
\label{wmass} (m_k^{\pm})^2 =   4 v^2 \sin^2{ \pi k \over N} +   (\rho^2
+ \eta^2) =   4 v^2 \sin^2{ \pi k \over N} +   |A|^2. \eeq The coupling
of the unbroken diagonal $U(1)$ at the scale $|A| = |\rho + i
\eta|$ is given by integrating out all $W^{\pm}$ since they are
all heavier than $A$; recall that we are working in a
normalization where $1/g_3^2$ multiplies the entire lagrangian
(\ref{kahler}, \ref{super}):
\beqa
  \label{u1coupling}
{1 \over
g_{3, D}^2(|A|)} &=& ~{N \over g_3^2} ~- ~
{1\over  2  \pi}~ \sum\limits_{k = -{N-1 \over 2}}^{N-1 \over 2} ~ {1
\over \sqrt{4   v^2 \sin^2
{\pi k \over N} +  {|A|^2}}}~ \nonumber \\
&\simeq&  ~{N \over g_3^2} ~-~ {N \over 4 \pi^2 v}~\int\limits_
{-{\pi\over 2}}^{\pi \over 2} {d x \over \sqrt{ \sin^2
x + {|A|^2 \over 4 v^2}}}  ~. \eeq This is, of course,  a regulated
version
of the continuum calculation, where (in the first equality, before
replacing the sum with an integral) we would have an infinite sum of
Kaluza-Klein tower
thresholds (\ref{3dmatching}) with $m_k = k^2 R^{-2} + |A|^2$.

The diagonal gauge coupling $g_{3, D}^2(|A|)$ of the deconstructed
theory   (\ref{u1coupling})    is identified, upon dimensional reduction,
  with  the running coupling of the continuum theory, see
\cite{Csaki:2001em}, \cite{Csaki:2001zx}.
The  4d running coupling  $1/g_4^2(|A|)$ is thus obtained   upon
multiplication of (\ref{u1coupling}) by $(2 \pi R)^{-1}$ and replacing
the sum with an integral  at large
$N$: \beq \label{u1coupling2} {1 \over g_4^2 (|A|)} \equiv {v \over
N g_{3, D}^2 (|A|)} = {v \over g_3^2}  - {1 \over 4 \pi^2} ~
\int\limits_{-{\pi\over 2}}^{\pi \over 2} {d x \over \sqrt{ \sin^2
x + {|A|^2 \over 4 v^2}}} ={v \over g_3^2}  + {1 \over 2
\pi^2}~  \ln {|A| \over 2 v}~. \eeq
The coefficient in front of the logarithm is   the one appropriate
  for the 4d $N=2$ pure $SU(2)$ Yang-Mills theory, which is usually written as
$b_0/(8\pi^2)$, with $b_0 = 4$. The continuum limit   in the
deconstructed  theory is then achieved
by taking (with $\Lambda$ fixed):
\beq \label{limit} N \rightarrow \infty, ~ v
\rightarrow \infty, ~{g_3^2\over v} \rightarrow 0,~ \Lambda^4 = v^4
~e^{- {8 \pi^2 v \over g_3^2}}, ~ \Lambda \ll v~.\eeq

Finally, to  compare with the supergravity calculation, we
note that the coupling on the probe brane, eqn.~(\ref{tadpole2}) with
$b_0 = 4$, is identical to the coupling of the diagonal 3d gauge group
(\ref{u1coupling}). Comparing (\ref{u1coupling}) with the tadpole
(\ref{tadpole2}) also gives a precise relation between   the integration
constant $C$  and the tree level coupling of the 3d theory: $C/2 =
1/g_3^2$.

\subsubsection{Supergravity and the $U(1)_R$ Anomaly}

It is clear from the supersymmetry of the problem that supergravity
should also yield the
  correct expression for the $U(1)_R$ anomaly in the continuum limit,
similar to the $N=2$ case
  discussed in section 3.1. Here, we briefly elaborate on this.

It is most convenient for the following discussion to work in
terms of the six dimensional theory one gets by orbifolding with
respect to $Z_r$ only. The closed string spectrum contains a
twisted R-R 1-form, which we denote $U$.\footnote{ This field is
untwisted with respect to $Z_N$, and therefore propagates in 6
dimensions. It is a combination of fields from the 3 dimensional
hyper- and vector multiplets. Like all closed string fields
relevant for us, it is in the $m=0$ Fourier mode with respect to
the $Z_r$ action.} The fractional D2 branes are magnetic sources
for this field (or electric sources for its dual
$\tilde{C}^{(3)}$). The D2 branes are defined as having: \beq
\int_{S^2} d U = \frac{b_0}{N} \eeq where $S^2$ is any two sphere
in the $3,8,9$ space surrounding only one 2-brane. The factor
$\frac{b_0}{N}$ is the R-R charge of a fractional brane, as
computed in section 3.2.

The tadpole of  the R-R twisted vector   is determined by the  $N$
sources at coordinates ${\bf X}_k$ (\ref{sources}), similar to
(\ref{tadpole0}):
\beq
\label{vector}
  (dU)^{ab} \sim  \frac{b_0}{N} ~\epsilon^{abc} ~
\partial_c ~ \sum_{k =1}^N {1 \over |{\bf X}  - {\bf X}_k| }~, \eeq
in an obvious notation.
Clearly, the field $U$ is the sum of the vector potentials for
magnetic monopoles located at ${\bf X}_k$. We will not need the general
expression for $U$, since near the probe brane and in the large-N
limit the r.h.s. of eqn.~(\ref{vector}) simplifies: as we already
showed in our discussion of the scalar tadpole  (\ref{tadpole2}),
there
   $ \sum_{k =1}^N {1 \over |X^a  - X^a_k| }  \sim \log |\rho + i \eta|$,
where
$\rho$ and $\eta$ are essentially $X^8$ and $X^9$ parametrizing
the tangent plane to the circle. Indexing $\rho$, $\eta$ by $i,j=
1,2$, we have from (\ref{vector}): \beq \label{twoform4} (d U)^{i
3} \sim  \epsilon^{i j 3}~ \partial_j \log |\rho + i \eta| ~. \eeq
Thus, near the probe brane the expression for the RR twisted
tadpole simplifies:
  \beq \label{twoform5} U_3 \sim {\rm
arg}(\rho + i \eta) ~. \eeq
The Chern-Simons coupling   to the twisted RR field  on the probe brane
world-volume $\int\limits_{{\cal V}_3} U \wedge F$ \cite{Douglas:1996sw}
leads, in the classical background (\ref{twoform5}), to:
\beq \label{coupling}
  \int d^3 x ~U_3~ \partial_m X^3 ~F_{p q} ~\epsilon^{m p q} \sim {\rm
arg} A \int d^3 x ~\partial_m X^3 ~F_{p q} ~\epsilon^{m p q}   \eeq
and gives  rise to the 4d anomaly (recall that $X^3$, the position of
the brane in the
compact direction,  becomes the third component of the
gauge field) exactly as in the $N=2$ case discussed in section 2.1.

\subsection{R-symmetry and the Wilson Term: a Geometrical Picture}

  Issues pertaining to fermion doublers and the anomalous  R-symmetry
away
  from the continuum limit are best visualized in the string theory
  embedding described above. Though it is simple enough to resolve
  such issues in the present context purely by field theoretical
  means, the stringy methods are expected to help in more complex
  situations.

  In our setup we have $N$ fractional branes which are distributed
  in an array in the $X^3,X^8$ plane, which is in turn embedded in
  the $X^3,X^8,X^9$ space. In the continuum limit the branes are
  distributed evenly on an approximate (for large $N$) circle, which is
to be considered as the
  direction conjugate (or T-dual) to the compact deconstructed
  direction.

  Let us consider the  open string  fields on the fractional branes. On
  each brane one has 3 massless scalar fields, corresponding to
  fluctuations of the brane in the space spanned by $X_3,X_8,X_9$.
   In the continuum limit two of those are to be considered as  a
complex  4-dimensional scalar
   field, and the remaining one as  a Wilson line of the 4-dimensional
gauge field around the
   compact direction.

   A priori this is  a local freedom, at each
   location of a fractional brane. This is represented in the field
theory language as
   a field redefinition freedom, which was used extensively in section
   2. Such choices differ, away from the continuum limit,
   by the value of the Wilson line turned on.

    We  choose to identify the complex scalar field as
   the fluctuations in the $X^8, X^9$ direction, away from the location
of a preferred
   brane, say the one with $k=0$.  In other words we identify $\rho,
\eta$, defined above, as the
   complex scalar field. This corresponds
   to turning off the   Wilson line,
   and as we see shortly,
   reproduces the Wilson R-symmetry breaking term.

    The identification of open string fields corresponds to a
    definition of the R-symmetry transformation away from the
    continuum limit. In general an R-symmetry transformation will
    be a rotation around some axis in the $X^3,X^8,X^9$ space. Our
    choice of fields corresponds to choosing the axis of rotation
    to be tangent to the circle of branes at the point $k=0$.

    It is clear then, that the locations of the $k\neq 0$ branes away
from the chosen tangent to the circle breaks the R-symmetry. Consider
for definiteness the tangent at the position of the
$k = 0$ brane, which, in the  coordinates of (\ref{sources}), is  at
$X^3 = 0$, $X^8 = v$. The continuum limit  R-symmetry is then an $SO(2)$
rotation in the $X^3,X^8, X^9$ space around the axis $X^8 = v, X^9 = 0$
(this axis is shown on Figure 2).

Recall now that the locations of the branes in the $X^3$--$X^8$ plane
correspond to   mass terms in the deconstructed theory. Having each
momentum mode arise from a brane at a different location in $X^8$
corresponds exactly to the (supersymmetric completion of)  Wilson
R-symmetry-breaking term.  The  masses of the adjoint supermultiplet due
to strings between the $0$-th and $k$-th brane are equal to the lengths
of the strings stretched between the two branes. The fermion mass itself
is complex and can be written as:
\beq
\label{geometric}
m_k =   e^{i {2 \pi k \over N}} ~v -  v = (X^8_k - v) + i ~X^3_k ~,
\eeq
where $X^{3 (8)}_k$ are the locations (\ref{sources}) of the
  $k$-th brane. The utility of writing the mass this way is to separate
the R-symmetry preserving (parallel to $X^3$) and R-symmetry breaking
(perpendicular to $X^3$) parts.
The dispersion relation that follows from (\ref{geometric}) is:
\beq \label{wilson3}
   v^{-2} \omega^2(k)   =  v^{-2}~|m_k|^2~=  \sin^2 {2 \pi k \over N} +
4  ~\sin^4 {\pi k \over N}~.
\eeq
Comparing (\ref{wilson3}) with (\ref{wilson1})
  identifies the Wilson term coefficient $r=1$. The value of the Wilson
term coefficient is thus of order the inverse lattice spacing, is fixed
by supersymmetry and is not a free parameter.

  \section*{Acknowledgements}

This work is supported by the National Science and Engineering
Research Council of Canada. E.P. acknowledges the hospitality and
support of the ``Braneworld and Supersymmetry" Workshop at the
Pacific Institute for Mathematical Sciences during the initial
stages of this work.

\end{document}